\documentclass[5p]{elsarticle}

\usepackage{amsmath,amssymb}
\usepackage{graphicx}
\usepackage{xcolor}
\usepackage{multirow}
\usepackage{bm}
\usepackage{nicefrac}
\usepackage{booktabs}
\usepackage{balance}
\usepackage{orcidlink}
\usepackage{makecell}
\usepackage{threeparttable}

\usepackage{hyperref}
\hypersetup{
  colorlinks=true,
  allcolors=blue
}

\begin{document}
\sloppy

\begin{frontmatter}

\title{Hydrogen uptake and hydride formation in \texorpdfstring{Al\textsubscript{\textit{x}}CoCrFeNi}{AlxCoCrFeNi} high-entropy alloys: First-principles, universal-potential, and experimental study}

\affiliation[inst1]{
organization={Interdisciplinary Centre for Advanced Materials Simulation (ICAMS), Ruhr-Universität Bochum},
addressline={Universitätsstr.\ 150},
postcode={44801},
city={Bochum},
country={Germany},
}

\affiliation[inst2]{
organization={Department for Computational Materials Design, Max-Planck-Institut for Sustainable Materials},
addressline={Max-Planck-Str.\ 1},
postcode={40237},
city={D\"usseldorf},
country={Germany},
}

\affiliation[inst3]{
organization={Institute for Materials Science, University of Stuttgart},
addressline={Pfaffenwaldring 55},
postcode={70569},
city={Stuttgart},
country={Germany},
}

\affiliation[inst4]{
organization={Deutsches Elektronen-Synchrotron DESY},
addressline={Notkestr.\ 85},
postcode={22607},
city={Hamburg},
country={Germany},
}

\affiliation[inst5]{
organization={Institute for Inorganic and Analytical Chemistry, Goethe University Frankfurt},
addressline={Max-von-Laue-Str.\ 7},
postcode={60438},
city={Frankfurt am Main},
country={Germany},
}

\affiliation[inst6]{
organization={Federal Institute for Materials Research and Testing (BAM)},
addressline={Richard-Willst\"atter-Str.\ 11},
postcode={12489},
city={Berlin},
country={Germany},
}

\affiliation[inst7]{
organization={Institute of Geology, Mineralogy and Geophysics, Faculty of Geosciences, Ruhr-Universität Bochum},
addressline={Universitätsstr.\ 150},
postcode={44801},
city={Bochum},
country={Germany},
}

\author[inst1,inst2]{Fritz K\"ormann \orcidlink{0000-0003-3050-6291}}
\ead{fritz.koermann@rub.de}

\author[inst3]{Yuji Ikeda \orcidlink{0000-0001-9176-3270}}

\author[inst4]{Konstantin Glazyrin \orcidlink{0000-0002-5296-9265}}

\author[inst5]{Maxim Bykov \orcidlink{0000-0003-0248-1728}}

\author[inst4]{Kristina Spektor \orcidlink{0000-0002-3267-9797}}

\author[inst4]{\\Shrikant Bhat}

\author[inst6]{Nikita Y. Gugin \orcidlink{0000-0002-7389-6528}}

\author[inst1]{Anton Bochkarev \orcidlink{0000-0001-7229-5758}}

\author[inst1]{Yury Lysogorskiy \orcidlink{0000-0003-4617-3188}}

\author[inst3]{\\Blazej Grabowski \orcidlink{0000-0003-4281-5665}}

\author[inst7]{Kirill V. Yusenko \orcidlink{0000-0002-8390-4288}}

\author[inst1]{{Ralf Drautz} \orcidlink{0000-0001-7101-8804}}

\begin{abstract}
Hydrogen uptake in complex multicomponent alloys, including high-entropy alloys (HEAs), governs both hydrogen storage capacity and resistance to hydrogen-induced degradation. We combine high-pressure experiments, density-functional theory (DFT), and a GRACE universal interatomic potential to investigate hydrogen absorption in Al$_{0.3}$CoCrFeNi and Al$_3$CoCrFeNi HEAs. In H$_2$ as a pressure-transmitting medium, the FCC Al$_{0.3}$CoCrFeNi alloy forms hydrides at ambient temperature above 3 GPa, whereas the Al-rich B2 Al$_3$CoCrFeNi alloy shows no evidence of hydride formation even upon heating at pressures up to 50 GPa. Experiments and calculations show that aluminum suppresses hydrogen uptake by increasing solution energies and destabilizing interstitial sites. The universal potential, employed in the calculations and pretrained on large DFT databases, closely reproduces DFT energetics and demonstrates transferability from the dilute limit to the hydride-forming regime. Simulations further disentangle the roles of local ordering, volume changes, composition, and crystal structure. Overall, our results indicate that hydrogen solubility in Al-containing HEAs is governed primarily by composition, with Al-driven B2 ordering as a strong secondary effect.
\end{abstract}


\end{frontmatter}

\section{Introduction}

\sloppy

Hydrogen uptake and hydride formation in high-entropy and other multicomponent alloys are central for both hydrogen storage \cite{marques2021review,LUO2024406} and the development of hydrogen-resistant structural materials \cite{li2022hydrogen}. 
In the dilute limit, the solution energy of hydrogen depends not only on chemical composition but also on the underlying crystal structure and the size of the available interstitial volume. 
Yet, in general, hydrogen solubility and transport are determined by a combination of geometric, electronic, and chemical factors, and do not usually follow a purely structural trend \cite{YIN2023105306}.

Aluminum-containing Al$_x$CoCrFeNi high-entropy alloys provide a prototypical system in which electronic structure, chemistry, and crystal structure are intrinsically linked. At low Al contents (up to $x \approx 0.3$), the alloy stabilizes in a single-phase face-centered cubic (FCC) structure, whereas increasing the Al concentration above $x \approx 0.9$ leads predominantly to a body-centered cubic (BCC) or primitive cubic B2 phase (CsCl-type) \cite{li2010effect,wang2014phases}. 
This change of structural stability from FCC to B2 is associated with pronounced differences in lattice volume, bonding character, and interstitial site geometry, all of which can influence hydrogen uptake.

Beyond structural considerations, other factors must be given due attention. Metallic aluminum exhibits a comparatively low solubility for hydrogen \cite{Wolverton2004}. Increasing the Al concentration may therefore reduce hydrogen uptake through chemical effects, while simultaneously altering phase stability and equilibrium volume, both of which strongly affect hydrogen accommodation. How these interacting chemical and structural effects counterbalance within the Al$_x$CoCrFeNi system has remained an open question.

High-pressure experiments in hydrogen as a pressure-transmitting medium provide a powerful tool to probe hydrogen uptake under extreme conditions, where elevated hydrogen chemical potentials can stabilize hydrides that are not stable at ambient pressure or temperature. The technique has been applied to several multicomponent alloys, revealing that some readily form hydrides and some remain essentially inert even when subjected to tens of gigapascals of hydrogen pressure \cite{glazyrin2024,glazyrin_natcomm_accepted}. While these measurements offer valuable macroscopic insight into hydrogen incorporation, they do not provide atomistic information on how local chemical and structural factors control the underlying energetics.

Density-functional theory (DFT) can effectively supplement the experimental studies, as it enables quantitative predictions of hydrogen solution energies and identifies favorable interstitial sites \cite{Wolverton2004}. Extensive DFT sampling of hydrogen configurations has been carried out for a few prototype high-entropy and multicomponent alloys \cite{qian2025hydrogen,wu2025hydrogen,moore2022hydrogen,ren2021hydrogen}. However, it remains computationally prohibitively expensive to apply such an approach systematically across many compositions and structures. Recently developed machine-learned interatomic foundational potentials, also known as "universal" potentials, offer a promising route to overcoming this limitation \cite{jacobs2025practical}. These potentials, including foundational potentials utilizing the Graph Atomic Cluster Expansion (GRACE) approach \cite{PhysRevX.14.021036,lysogorskiy2025graphatomicclusterexpansion}, have demonstrated broad transferability across chemically complex metallic systems, enabling DFT-comparable accuracy at a fraction of the computational cost \cite{zhang2025lattice}. However, their capability to describe hydrogen energetics, both in the dilute-interstitial limit and in concentrated hydrides, remains unchallenged for chemically and compositionally complex, disordered alloys.

In this work, we investigate hydrogen uptake in two representative single-phase alloys: FCC Al$_{0.3}$CoCrFeNi and B2 (CsCl-type) Al$_3$CoCrFeNi. 
We use DFT calculations and GRACE foundational potentials to quantify hydrogen solution energies, assess site stability, and evaluate hydride-formation tendencies, which are compared with high-pressure experiments. This complementary approach allows us to probe and to disentangle the respective roles of composition, lattice volume, and crystal structure in controlling hydrogen uptake. We find that hydrogen affinity in these alloys is primarily governed by the aluminum content and associated volume effects. In contrast, differences between FCC and BCC/B2 crystal structures has a secondary role once composition and volume are fixed.

\section{Methods}

\subsection{Computational Details}\label{compDetails}

Spin-polarized DFT calculations were performed using the Vienna \textit{Ab initio} Simulation Package (VASP, version~5.4.4)~\cite{kresse1993,kresse1994,kresse1996}, employing the projector-augmented wave (PAW) method~\cite{blochl1994}. Initial magnetic moments were assigned to the transition-metal elements (Co, Fe, Ni, and Cr), while Al was initialized with zero magnetic moment. The exchange–correlation energy was described using the Perdew–Burke–Ernzerhof (PBE) generalized gradient approximation (GGA)~\cite{perdew1996}. All calculations used an energy cutoff of 300~eV, Gaussian smearing of 0.2~eV, and $3\times3\times3$ $k$-point meshes and $6\times6\times6$ $k$-point meshes for the 216-atom and 64-atom supercells, respectively. Structural relaxations allowed internal atomic positions to relax while keeping the cell shape fixed (\texttt{ISIF=2}). Chemically disordered alloy configurations were represented by special quasi-random structures (SQSs)~\cite{zunger1990special} generated with the \textsc{icet} package~\cite{aangqvist2019icet}. 
The B2 structure was generated by placing all Al atoms on the corner sites, while all remaining atoms were distributed randomly across the unoccupied lattice positions. 

To model the two compositions Al$_3$CoCrFeNi and Al$_{0.3}$CoCrFeNi, we employed the following setup to match the experimental compositions: The 64-atom supercells correspond to Al$_{28}$Co$_9$Cr$_9$Fe$_9$Ni$_9$ (Al$_{3.11}$CoCrFeNi) and Al$_4$Co$_{15}$Cr$_{15}$Fe$_{15}$Ni$_{15}$ (Al$_{0.27}$CoCrFeNi), while the 216-atom supercells correspond to Al$_{92}$Co$_{31}$Cr$_{31}$Fe$_{31}$Ni$_{31}$ (Al$_{2.97}$CoCrFeNi) and Al$_{16}$Co$_{50}$Cr$_{50}$Fe$_{50}$Ni$_{50}$ (Al$_{0.32}$CoCrFeNi).

The ionic convergence criteria (\texttt{EDIFFG}) were set to ten times the electronic criteria (\texttt{EDIFF}). For the 64-atom cells, we used  \texttt{EDIFF} = $1\times10^{-5}$, corresponding to an ionic convergence better than $0.1$\,meV per supercell, while for the 216-atom cells  \texttt{EDIFF} = $1\times10^{-2}$ was sufficient to ensure convergence better than $0.5$\,meV per atom. These different targets reflect that the 216-atom cells were compared on a per-atom energy basis (pure alloys and hydrides). In contrast, the 64-atom cells were used to evaluate the performance of the foundational potential in predicting solution energies on a per-supercell basis.

The H$_2$ molecule was modeled in a 20~\AA{} cubic box, employing the same
300~eV plane-wave cutoff as used for the hydride and H-solution calculations, and
sampled at the $\Gamma$-centered point.

All alloy configurations were evaluated using the GRACE framework~\cite{PhysRevX.14.021036} together with the recently introduced universal foundational potentials~\cite{lysogorskiy2025graphatomicclusterexpansion}. We employed the \texttt{GRACE\_2L\_SMAX\_OMAT\_large} model, which was trained on a subset of the OMat24 dataset \cite{barroso2024open} (excluding all calculations using a +U-correction) together with more than one million structures generated via the maximum-entropy principle \cite{pa2025information}, also computed without +U-correction. Details of the potential training procedure are described in Ref.~\cite{maxEntropy_foundation_submitted}. For each supercell size, all alloy variants and all octahedral and tetrahedral interstitial sites were evaluated using the Atomic Simulation Environment (ASE)~\cite{larsen2017atomic} utilizing the BFGS optimizer with forces converged to below 0.01~eV/\AA.

\subsection{Experimental Details}

The two alloys, FCC Al$_{0.3}$CoCrFeNi and B2  Al$_{3}$CoCrFeNi, were prepared by induction melting of stoichiometric mixtures of high-purity elemental powders \cite{Yusenko2018}. Melting was performed in $h$-BN crucibles placed inside an Ar-filled glove box to ensure an inert processing environment. Complete melting of both alloy compositions was achieved at temperatures exceeding 1580 K. 
Following a holding time of 1–2 minutes at the melting temperature, the melts were allowed to cool naturally to ambient conditions. 
To promote chemical homogeneity, each alloy ingot was subsequently re-melted three additional times under identical conditions. 
A detailed description of the physicochemical characteristics and phase constitution of these alloys can also be found in Ref.~\cite{Yusenko2018}.

Room-temperature compressibility data for both alloys were measured in Ne and H$_2$ as pressure-transmitting media using diamond anvil cells (DACs) at the P02.2 beamline of PETRA-III, DESY \cite{Liermann2015}. Measurements were conducted up to 55~GPa at room temperature, using an X-ray wavelength of $\lambda = 0.2907$~Å and a Perkin Elmer XRD1621 detector (beam size 3(v)~$\times$~8(h)~$\mu$m$^2$, FWHM).  The samples were loaded into symmetric DACs equipped with 300~$\mu$m Boehler–Almax anvils. Ruby and gold powder placed close to the sample as pressure calibrants \cite{Fei2007}. Ruby was used for the pre-compression purposes and the pressures shown in the figures below were extracted from Au EOS for consistency. Diffraction images were integrated with DIOPTAS \cite{Prescher2015}. Unit-cell parameters, background, and profile functions were refined using TOPAS \cite{Coelho2018}, and the obtained pressure-volume data were fitted using EoS-Fit 5.2 \cite{GonzalezPlatas2016}.

High-temperature compression data for the FCC Al$_{0.3}$CoCrFeNi alloy in a H$_2$ fluid were collected using the Hall-type Aster–15 six-ram, large-volume multi-anvil press (LVP, MAVO LPQ6-1500-100) at the P61B energy-dispersive beamline of PETRA-III, DESY \cite{Farla2022}. Approximately 5~mg of alloy powder was pressed into a pellet (1.2~mm outer diameter, 0.8~mm height) and sealed inside a NaCl capsule (3~mm outer diameter, 3.55~mm height) together with two NH$_3$BH$_3$ pellets of identical size. NH$_3$BH$_3$ (Sigma Aldrich, 97\%) served as the hydrogen source, supplying a $\sim$2.5-fold molar excess of H relative to the metal content. Upon decomposition induced by heating at high pressure, BH$_3$NH$_3$ yields inert BN as the solid residue \cite{Nylen2009}. Complete capsule preparation was carried out in an Ar-filled glove box. The samples were compressed using 14/7 multi-anvil assemblies \cite{Vekilova2023}. Pressure was determined from the NaCl equation of state \cite{matsui2012simultaneous}, and the temperature was calibrated using power–temperature curves established from offline in situ LVP runs employing Type-C thermocouples.

The sample assembly was compressed to 10~GPa and then gradually heated to 720~$^\circ$C to decompose NH$_3$BH$_3$ and generate H$_2$. 
After equilibration, it was cooled to room temperature and decompressed to ambient pressure. 
PXRD data were collected continuously during compression, heating, cooling, and decompression using the high-energy white X-ray beam as a probe with a high-purity Ge energy-dispersive detector equipped with a collimator--slit system positioned at the diffracted angle $2\theta = 4.9621^\circ$. Energy-dispersive diffraction patterns were analyzed using the PDIndexer software \cite{Seto2010}.

\section{Results and Discussion}

\subsection{Validation of the foundational potential}\label{sec:validation}

\begin{figure}[t]
            \includegraphics[width=\linewidth]{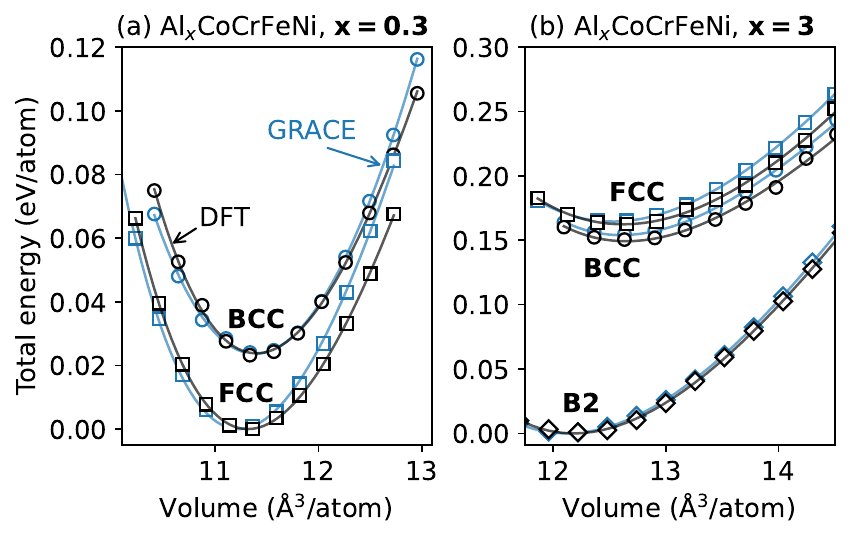}
    \caption{Energy–volume curves for Al$_x$CoCrFeNi with (a) $x=0.3$ and (b) $x=3.0$ for 216-atom supercell models. Predictions using the GRACE foundational potential (blue) and DFT calculations (black) are presented for the BCC and FCC phases. For Al$_3$CoCrFeNi, the B2 phase is also included. Symbols denote the calculated data, and lines show the fits using the Vinet EOS.
    \label{fig:eos}}
\end{figure}

We first examine the phase stability of the hydrogen-free Al$_{0.3}$CoCrFeNi and Al$_{3}$CoCrFeNi parent alloys. We computed total energies for the FCC and BCC structures using 216-atom supercells with both the GRACE foundational potential and explicit DFT calculations. For each composition and structure, we used identical atomic configurations in the DFT and GRACE calculations to enable a direct, method-to-method comparison. 
In addition to the random BCC structure, we also constructed an ordered B2 phase for Al$_3$CoCrFeNi, which has been reported experimentally in the literature~\cite{wang2014phases,smekhova2022driven}. The resulting energy--volume curves are shown in Fig.~\ref{fig:eos}. The experimentally established phase stability---FCC for Al$_{0.3}$CoCrFeNi and B2 for Al$_3$CoCrFeNi---is well reproduced. The equation-of-state (EOS) parameters are summarized in Table~\ref{tab:eos}. For GRACE and DFT, the equilibrium energy differences, volumes, and bulk moduli were estimated by fitting the calculated energy–volume data to the Vinet EOS~\cite{PhysRevB.35.1945}, while the experimental values correspond to EOS parameters fitted from high-pressure X-ray diffraction measurements in Ne as a pressure-transmitting medium \cite{Yusenko2018}. Overall, GRACE and DFT yield consistent equilibrium properties that are in reasonable agreement with experiment. For FCC Al$_{0.3}$CoCrFeNi, the results are also consistent with ambient-pressure lattice parameters of the as-cast alloy measured by XRD \cite{zhao2023investigation}. 

\begin{table}[t]
\centering
\caption{Equilibrium EOS parameters from GRACE (GRA), DFT, and experiment (Exp).
$\Delta E_\mathrm{eq}$ is referenced to the lowest-energy phase within each composition (bold).}
\label{tab:eos}

\begin{threeparttable}
\footnotesize
\setlength{\tabcolsep}{3pt}
\renewcommand{\arraystretch}{1.1}

\begin{tabular}{l | cc | ccc}
\hline
 & \multicolumn{2}{c|}{\textbf{Al$_{0.3}$}-HEA} & \multicolumn{3}{c}{\textbf{Al$_3$}-HEA} \\
\cline{2-6}
 & BCC & FCC & FCC & BCC & B2 \\
\hline
$\Delta E_\mathrm{eq}$ (meV/at.), GRA & 24 & \textbf{0} & 164 & 154 & \textbf{0} \\
$\Delta E_\mathrm{eq}$ (meV/at.), DFT & 24 & \textbf{0} & 150 & 149 & \textbf{0} \\\hline
$V_\mathrm{eq}$ (\AA$^3$/at.), GRA    & 11.39 & \textbf{11.23} & 12.53 & 12.60 & \textbf{12.12} \\
$V_\mathrm{eq}$ (\AA$^3$/at.), DFT    & 11.40 & \textbf{11.32} & 12.62 & 12.65 & \textbf{12.17} \\
$V_\mathrm{eq}$ (\AA$^3$/at.), Exp.    & --- & \textbf{11.56}\tnote{a} & --- & --- & \textbf{11.97}\tnote{a} \\
&     & 11.62\tnote{b} &     &     &  \\
\hline
$B_\mathrm{eq}$ (GPa), GRA            & 154 & \textbf{176} & 135 & 146 & \textbf{151} \\
$B_\mathrm{eq}$ (GPa), DFT            & 161 & \textbf{161} & 127 & 137 & \textbf{151} \\
$B_\mathrm{eq}$ (GPa), Exp.            & --- & \textbf{201}\tnote{a} & --- & --- & \textbf{197}\tnote{a} \\\hline
$B'_\mathrm{eq}$, GRA                 & 2 & \textbf{6} & 5 & 8 & \textbf{5} \\
$B'_\mathrm{eq}$, DFT                 & 5 & \textbf{6} & 4 & 7 & \textbf{5} \\
$B'_\mathrm{eq}$, Exp.                 & --- & \textbf{4}\tnote{a} & --- & --- & \textbf{4}\tnote{a} \\
\hline
\end{tabular}

\begin{tablenotes}\footnotesize
\item[a] Experimental EOS parameters from Ref.~\cite{Yusenko2018}.
\item[b] Lattice parameter from Ref.~\cite{zhao2023investigation}.
\end{tablenotes}
\end{threeparttable}
\end{table}

\begin{figure}[th!]
\includegraphics[width=0.9\linewidth]{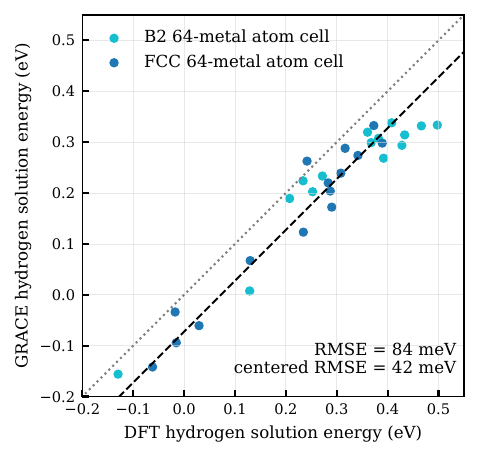}
\caption{
Comparison of hydrogen solution energies obtained from DFT and the GRACE foundational potential for FCC Al$_{0.3}$CoCrFeNi and the B2-ordered Al$_3$CoCrFeNi using 64-atom supercells. Each point corresponds to a unique interstitial configuration. The black dashed line represents the correlation with a mean signed difference of $\Delta\mu = -72$~meV, as determined by linear regression, indicating a systematic difference in the hydrogen chemical potential.\label{fig:Esol_DFT_GRACE} }
\end{figure}

We next validate the performance of the foundational potential for hydrogen solution energies,
\begin{equation}
\Delta E_\mathrm{sol} = E(\mathrm{alloy+H}) - E(\mathrm{alloy}) - \mu_\mathrm{H},
\end{equation}
where $E(\mathrm{alloy+H})$ and $E(\mathrm{alloy})$ denote the total energies of the alloys per simulation cell with and without the hydrogen atom, respectively, and $\mu_\mathrm{H}=\tfrac{1}{2}E_{\mathrm{H}_2}$ is the hydrogen chemical potential. In contrast to the non-hydrogenated EOS energies, which are compared as per-atom quantities, the solution energies are evaluated as differences per supercell. The latter comparison provides a more rigorous validation. Specifically, we calculated the energies in 64-atom supercells, as detailed in Section \ref{compDetails}. Smaller supercells enable more accurate DFT parameters, which are essential for achieving high precision per supercell. We restricted the analysis to the experimentally stable phases, namely FCC Al$_{0.3}$CoCrFeNi and B2 Al$_3$CoCrFeNi. All calculations were initialized with atoms placed on ideal FCC or BCC-based B2 lattice sites and the interstitial hydrogen atom located at the corresponding ideal interstitial site, followed by complete atomic relaxation in both methods. As shown in Fig.~\ref{fig:Esol_DFT_GRACE}, the two methods exhibit a strong linear correlation. A mean signed difference of approximately 72~meV per H atom, obtained from a constrained regression with fixed unit slope, reflects a systematic shift in the reference hydrogen chemical potential associated with the H$_2$ molecule in the foundational potential employed, likely because the foundational potential does not explicitly include H$_2$ energetics. The corresponding centered RMSE, obtained by subtracting the systematic shift, is 42 meV per H atom, which translates to an accuracy of well below one meV per atom. 
More importantly, our results suggest that the qualitative trends and relative stability of hydrogen sites are accurately captured.

To complement the analysis of the just-discussed dilute-limit solution energies and to assess the stability of highly hydrogenated phases, we further examine two idealized 1:1 metal--hydrogen hydrides, corresponding to a concentration of one hydrogen atom per metal atom (H/M = 1). 
For both systems, we employed the same 216 metal-atom supercells as used for the non-hydrogenated EOS calculations discussed above and derived their energetics from newly computed energy--volume EOSs. The FCC \(\mathrm{Al_{0.3}CoCrFeNi}\) hydride with H/M = 1 was constructed by filling all octahedral interstitial sites. In contrast, for B2 \(\mathrm{Al_3CoCrFeNi}\), we populated the most favorable interstitial positions identified by GRACE (predominantly tetrahedral sites) to construct the hydrogenated structure. Both hydrides were assessed using the GRACE foundational potential and explicit DFT calculations. As summarized in Table~\ref{tab:eos_hydrides}, hydriding leads to effective hydrogen atomic volumes of approximately 2\,\text{\AA}\textsuperscript{3} per atom in the FCC alloy and nearly 4\,\text{\AA}\textsuperscript{3} per atom in the B2 alloy. The substantially larger effective hydrogen volume in the B2 phase indicates a greater volumetric penalty associated with hydrogen incorporation than in the FCC alloy. The respective volume changes are in reasonable agreement between the GRACE foundational potential and the explicit DFT
calculations.

When comparing \textit{absolute} hydride formation energies, the deviations between the GRACE results and DFT are larger: \textminus60~meV and \textminus61~meV per metal atom for the FCC hydride and B2 hydride, respectively. The nearly identical magnitudes of these deviations indicate a systematic shift, which we again attribute to the hydrogen reference chemical potential. As this offset is essentially constant, it does not affect the relative stability between the two hydrides. A clear formation energy favoring the FCC hydride is observed compared to the B2 hydride, with a predicted difference in formation energies of 277~meV using the foundational potential and 278~meV from the explicit DFT calculations. 

\begin{table}[ht]
\centering
\caption{Hydride formation energies ($E_\mathrm{form}$) and the associated volume increases ($\Delta V_\mathrm{H}$) for hydrides with H/M = 1, obtained with GRACE and DFT.
Hydrogen atoms occupy interstitial sites as described in the main text, yielding a 1:1 metal--hydrogen ratio.
$\Delta V_\mathrm{H}$ corresponds to the effective hydrogen volume.
$\Delta E_\mathrm{form}^{\mathrm{B2-FCC}}$ denotes the formation-energy difference between the FCC $\mathrm{Al_{0.3}CoCrFeNi}$- and B2 $\mathrm{Al_3CoCrFeNi}$-based hydrides. Values in brackets are obtained from non-spin-polarized calculations.}
\label{tab:eos_hydrides}

\footnotesize
\setlength{\tabcolsep}{4pt}
\renewcommand{\arraystretch}{1.15}

\begin{tabular}{lcccc}
\hline
 & \multicolumn{2}{c}{$E_\mathrm{form}$ (meV/metal)}
 & \multicolumn{2}{c}{$\Delta V_\mathrm{H}$ (\AA$^3$/H)} \\
Hydrogenated phase & GRACE & DFT & GRACE & DFT \\
\hline
FCC $\mathrm{Al_{0.3}CoCrFeNi}$ & $-53$ & 7 (4)     & 2.04 & 2.11 \\
B2  $\mathrm{Al_3CoCrFeNi}$     & 224   & 285 (283) & 3.69 & 4.02 \\
\hline
$\Delta E_\mathrm{form}^{\mathrm{B2-FCC}}$\rule[-1.3ex]{0pt}{3.9ex}
 & 277 & 278 (279) & \multicolumn{2}{c}{--} \\
\hline
\end{tabular}
\normalsize
\end{table}

\subsection{Hydrogen uptake from theory and experiment}

\begin{figure*}
    \centering
    \includegraphics[width=1\linewidth]{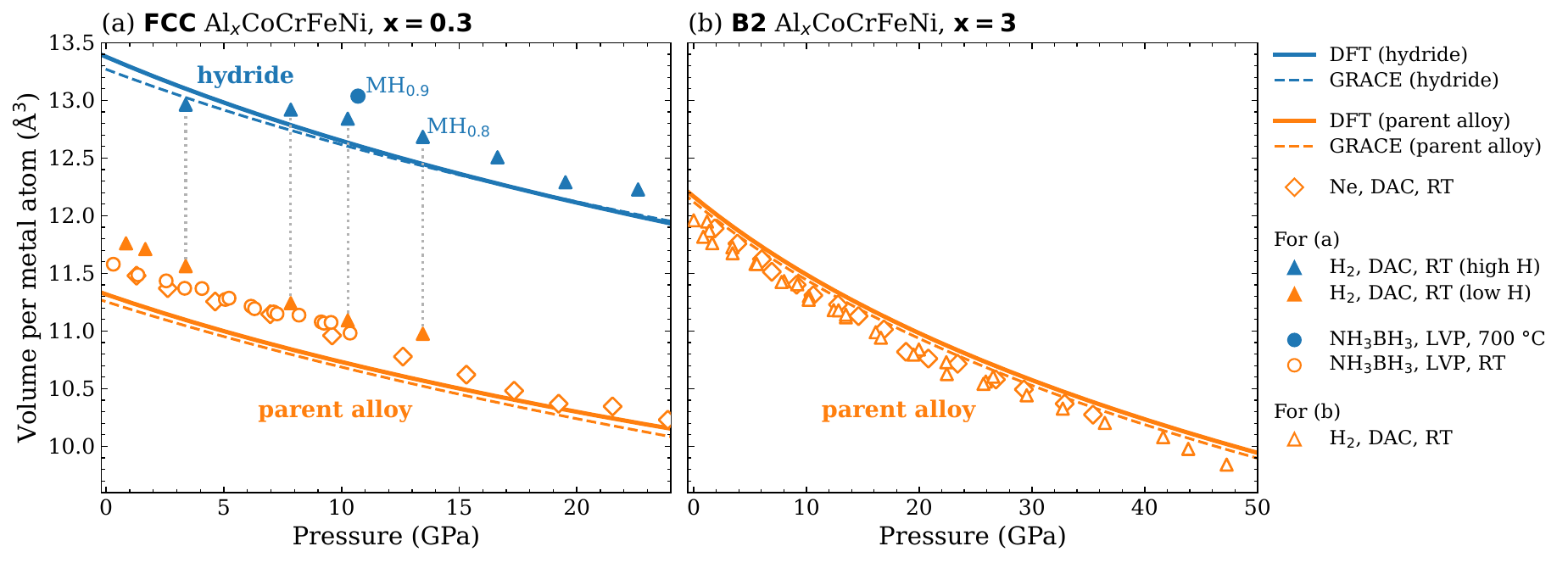}
    \caption{(a) Compressibility data obtained from DAC and LVP experiments for FCC~Al$_{0.3}$CoCrFeNi. The samples collected in DACs were surrounded by either H2 or Ne pressure-transmitting medium. 
Symbols denote experimental measurements. The dotted vertical lines indicate the pressure range in which two FCC phases, with lower and higher volumes, respectively, coexist. The data associated with solid triangles was collected in a DAC on compression. LVP data (open circles) were collected at room temperature during compression and then at high temperature (closed circle). Solid and dashed lines show pressure–volume curves of the hydrogen-free parent alloy and of the ideal hydride with H/M = 1, respectively, obtained from DFT and GRACE calculations for comparison.
(b) Corresponding DAC measurements for B2~Al$_3$CoCrFeNi in H$_2$ and Ne, demonstrating the absence of hydride formation up to 50\,GPa. GRACE and DFT calculated EOS curves for the hydrogen-free parent alloys are shown for reference.}
    \label{fig:experiments}
\end{figure*}

The calculations performed above consistently show that hydrogen incorporation is energetically favorable in FCC~Al$_{0.3}$CoCrFeNi, whereas it is unfavorable in B2~Al$_3$CoCrFeNi. This qualitative trend is reproduced by both DFT and the GRACE foundational potential, with hydrogen insertion in the B2 phase associated with an energetic penalty of approximately 0.3~eV.

To examine whether this predicted structural dependence of hydrogen stability is reflected experimentally, we performed complementary high-pressure experiments, summarized in Fig.~\ref{fig:experiments}. High pressures impose a large hydrogen chemical potential and therefore promote hydride formation \cite{glazyrin2024}. 
We do not attempt a quantitative mapping between pressure–temperature conditions and the hydrogen chemical potential. Instead, we use the volume-pressure relations $V(p)$ as a direct experimental indicator of hydrogen uptake, where discontinuities or pronounced growth in $V(p)$ signal hydride formation. 

For FCC Al$_{0.3}$CoCrFeNi, Figure~\ref{fig:experiments}(a) clearly shows signs of hydrogen absorption. When the alloy is compressed at room temperature above approximately 3\,GPa, a noticeable increase in atomic volume occurs, indicating the formation of a hydrogen-containing phase. To estimate the hydrogen content, we used the DFT-computed effective hydrogen volume (cf.~Table~\ref{tab:eos_hydrides}), which is close to the empirical value of about 2.1\,\text{\AA}\textsuperscript{3} per H atom commonly employed for similar FCC alloys~\cite{glazyrin_natcomm_accepted,somenkov1987crystal,fukai2005metal}. A hydrogenated phase with a metal-to-hydrogen ratio of about MH$_{0.8}$ (denoted as high~H) is observed. High-temperature high-pressure experiment in the LVP pushes the system toward equilibration. As a result, a higher hydrogen content, with approximately MH$_{0.9}$, has been observed. For comparison, the computed pressure–volume curves derived from DFT and the GRACE foundational potential for both the hydrogen-free parent alloy and the corresponding hydride with   H/M = 1 are presented in Figure~\ref{fig:experiments}(a) as thick solid and dashed lines, respectively. While both theoretical models slightly underestimate the absolute volumes, they accurately reflect the experimental trends and capture the extent of volume expansion upon hydrogenation.

In contrast, Fig.~\ref{fig:experiments}(b) shows that the B2-ordered Al$_3$CoCrFeNi alloy exhibits no evidence of hydride formation across the entire pressure range studied at ambient temperature. The B2 alloy does not show any detectable growth of its atomic volume in the presence of hydrogen, compared with the pristine compressibility experiment in an inert pressure-transmitting medium (Ne or He) \cite{Yusenko2018}. Our GRACE and DFT calculations, which predict an energetic penalty of approximately 0.3 eV for hydrogen incorporation in the B2 phase, support this finding. The alloy exhibits a smooth, continuous compression curve up to 50~GPa, without any detectable volume anomalies indicative of hydrogen uptake or phase transformations. 

Moreover, the calculated EOS for the hydrogen-free B2 parent alloy, obtained from both GRACE and DFT, aligns well with the experimentally measured $V(p)$, corroborating the absence of hydrogen incorporation under pressure. Even after heating to 150~°C, no significant deviation from the hydrogen-free compression behavior is observed (see Supplemental Material). This clear qualitative difference underscores the much lower hydrogen affinity of the B2 phase compared to its FCC counterpart.

The systematic underestimation of the experimental equilibrium volume for the FCC Al$_{0.3}$CoCrFeNi alloy and the hydride, as calculated using PBE, is consistent with previous findings reported for the Al-free CoCrFeNi alloy~\cite{WANG2019468}. This behavior aligns with known trends of PBE for transition metals, where the equilibrium volumes of Fe, Co, and Ni are underestimated even for pure elements~\cite{PhysRevB.87.214102}. In contrast, PBE results in equilibrium volumes for Al that are larger than those observed experimentally~\cite{Lejaeghere_2006,PhysRevB.79.085104}. The closer agreement with the experimental results reported here for the Al-rich B2 Al$_3$CoCrFeNi alloy can therefore be qualitatively interpreted as a partial compensation between these opposing trends for the individual elements.

\subsection{\texorpdfstring{Chemical and structural factors governing hydrogen uptake in Al\textsubscript{\textit{x}}CoCrFeNi}{Chemical and structural factors governing hydrogen uptake in AlxCoCrFeNi}}

\begin{figure}[th!]
\includegraphics[width=\linewidth]{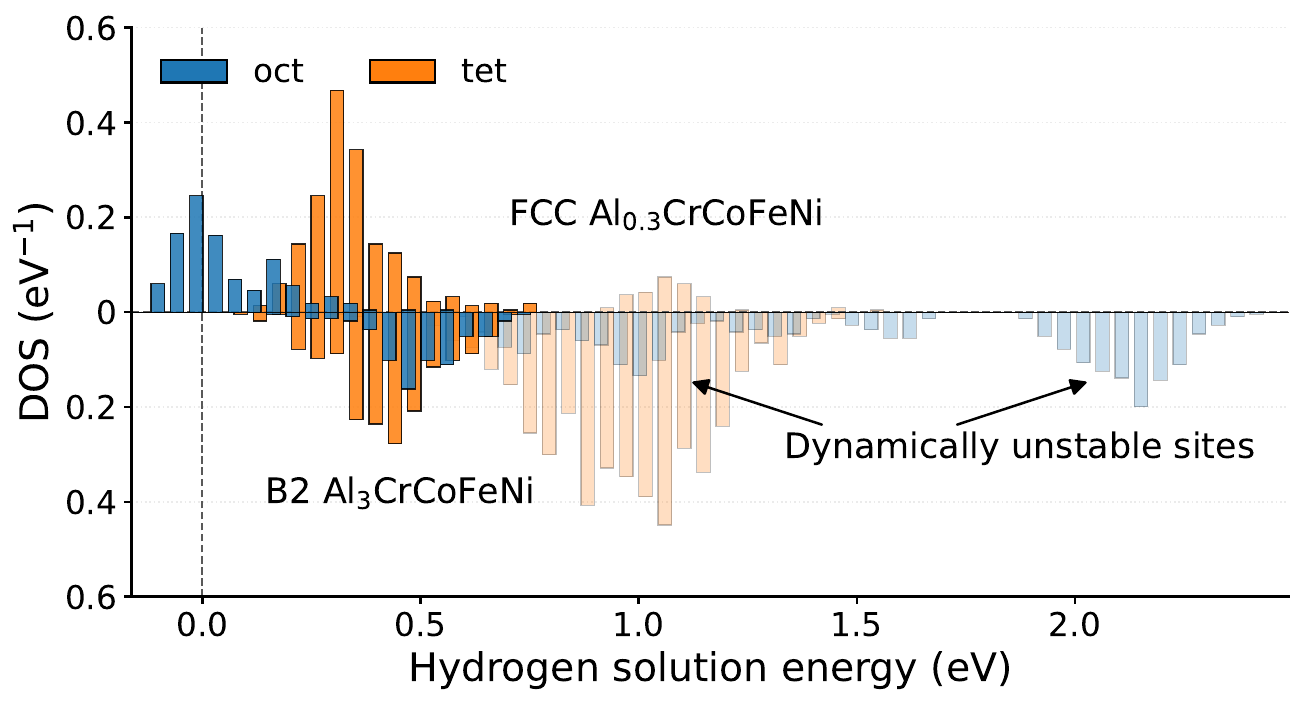}
   \caption{
Histograms of hydrogen solution energies for interstitial octahedral and tetrahedral sites in 
FCC~Al$_{0.3}$CoCrFeNi (upper part) and B2~Al$_3$CoCrFeNi (lower part) obtained using the GRACE foundational potential. 
The bar height represents the number of sites per metal atom.
Semi-transparent bars indicate dynamically unstable configurations in which the hydrogen atom relaxes 
from its initial site to a more stable one.\label{fig:Esol}
}
\end{figure}

To understand the atomistic origin of the difference in hydrogen affinity between FCC~Al$_{0.3}$CoCrFeNi and B2~Al$_3$CoCrFeNi, we performed a comprehensive screening of hydrogen solution energies across a wide range of structural, compositional, and volumetric scenarios. Such an extensive sampling would be computationally prohibitive using explicit DFT alone. We thus utilized the efficient GRACE foundational potential, which we carefully confirmed reliably reproduces DFT energetics for both hydrogen-free alloys and hydrogen-containing reference structures in Sec.~\ref{sec:validation}.

\begin{figure*}[t]
\includegraphics[width=\linewidth]{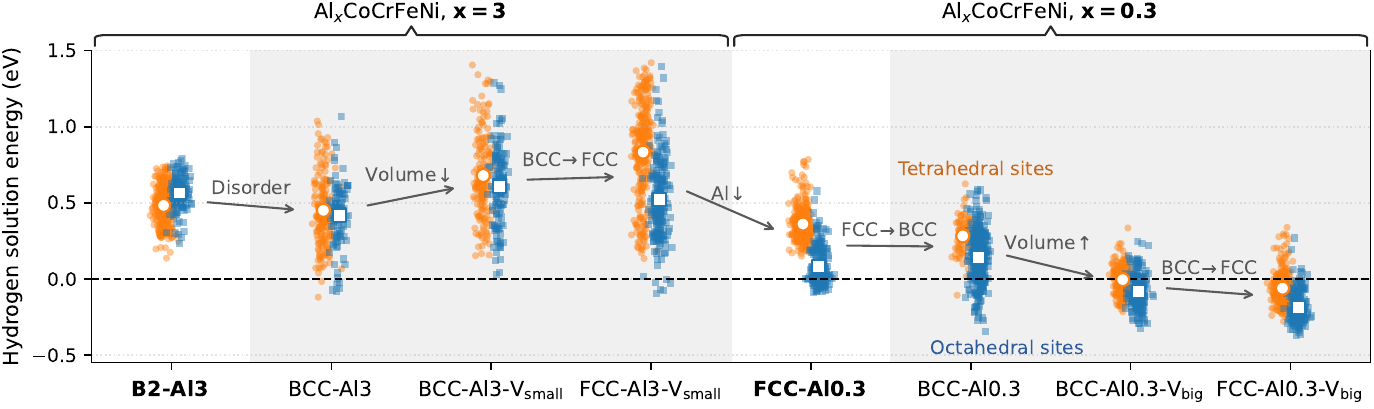}
   \caption{
Solution energies of hydrogen for tetrahedral (orange) and octahedral (blue) interstitial sites across different structures, compositions, and volumes of the Al-containing HEAs, obtained using the GRACE foundational potential. Each state is shown as a distribution of site-resolved energies, with mean values indicated by outlined markers: blue squares for tetrahedral sites and orange circles for octahedral sites. Arrows denote the corresponding change in chemical order, volume, structure, or Al content. The black-dashed line marks zero-solution energy. 
The stable reference states (B2~Al$_3$CoCrFeNi and FCC~Al$_{0.3}$CoCrFeNi) are highlighted.\label{fig:master_solution}
}
\end{figure*}

To assess the hydrogen affinity, we considered the hydrogen solution energies for the 216-atom supercells. We evaluated all interstitial sites in the FCC, BCC, and B2 phases of Al$_{0.3}$CoCrFeNi and Al$_3$CoCrFeNi. Zero-point vibrational contributions were neglected, as they are expected to largely cancel in hydrogen solution and in formation energies when comparing different sites, structures, and compositions, and are therefore not expected to affect the relative trends discussed here. In the 216-atom supercells considered, the BCC lattice hosts 648 octahedral and 1296 tetrahedral sites, while the FCC lattice has 216 octahedral and 432 tetrahedral sites, yielding a total of 5184 interstitial configurations across both alloys. 

We first analyze the hydrogen solution energies obtained for the stable phases of both alloys, summarized in Fig.~\ref{fig:Esol}. The upper panel shows all interstitial configurations in FCC Al$_{0.3}$CoCrFeNi, while the lower panel presents those in B2 Al$_3$CoCrFeNi. Overall, FCC Al$_{0.3}$CoCrFeNi exhibits markedly lower hydrogen solution energies, by roughly 0.3~eV per H atom, compared to B2 Al$_3$CoCrFeNi, which is consistent with the difference in hydride formation energies, cf. Table~\ref{tab:eos_hydrides}. For B2 Al$_3$CoCrFeNi, many interstitial configurations are dynamically unstable, with hydrogen relaxing away from the initially imposed prototype position into a different, energetically more favorable interstitial environment, most frequently of tetrahedral character. These unstable configurations were identified by comparing the relaxed hydrogen positions to a set of ideal interstitial prototypes and assigning each to the nearest interstitial environment. To provide a complete overview of the available interstitial landscape, such unstable configurations are shown using their unrelaxed energies (faded bars in Fig.~\ref{fig:Esol}). This comparison clearly indicates that FCC Al$_{0.3}$CoCrFeNi has significant stable (negative) solution energies, whereas B2 Al$_3$CoCrFeNi has very few low-energy sites, making it thus significantly more resistant to hydrogen uptake.

We next examine how composition, chemical order, and volume determine the hydrogen solution energies in these alloys. In particular, the substantial volume expansion associated with increasing Al content, the intrinsically low affinity of Al for hydrogen, and changes in chemical order, e.g., from Al-ordered in B2 to disordered in FCC, represent three competing effects. To disentangle their individual contributions, we constructed a sequence of intermediate states connecting the two stable reference alloys, B2--Al$_3$CoCrFeNi and FCC--Al$_{0.3}$CoCrFeNi. 

For each intermediate state, all available tetrahedral and octahedral interstitial sites were systematically evaluated. To construct the distributions shown in Fig.~\ref{fig:master_solution}, we focused on the dynamically stable sites, which simplifies the visualization while still capturing all relevant energetic trends.

The first transition in Fig.~\ref{fig:master_solution}, from the ordered B2 Al$_3$CoCrFeNi to chemically disordered BCC Al$_3$CoCrFeNi, clarifies the impact of chemical order. The disordered alloy exhibits a broader distribution with several lower-energy sites, reflecting the presence of locally Al-poor environments that occur much more frequently in the random alloy as compared to the B2-ordered structure. While low-energy hydrogen solution sites are statistically rare in the B2-ordered alloy, they are not strictly forbidden. According to our calculations, for example, one such locally Al-poor configuration is observed in the 64-atom B2 cell (light-blue points in Fig.~\ref{fig:Esol_DFT_GRACE}).

In the next step, the BCC Al$_3$CoCrFeNi alloy is compressed in our simulations from its large equilibrium volume (12.13~\AA$^3$/atom, cf.~Table~\ref{tab:eos}) to a smaller volume of $11.25$~\AA$^3$/atom to mimic the volume of the Al$_{0.3}$CoCrFeNi composition (Table~\ref{tab:eos}). This reduction, which mimics the volume effect of decreasing Al content, increases the solution energies and illustrates that the larger equilibrium lattice volume associated with Al-rich compositions actually facilitates hydrogen uptake. At this constrained volume, a structural change from BCC to FCC stabilizes several octahedral sites, but the overall solution energies remain comparatively high. In the subsequent step, the transition from the volume-compressed FCC Al$_3$CoCrFeNi to FCC Al$_{0.3}$CoCrFeNi stabilizes the majority of octahedral sites for hydrohen uptake, reflecting the much higher availability of Al-poor configurations in the Al$_{0.3}$CoCrFeNi alloy.

The sequence is then mirrored for the Al$_{0.3}$CoCrFeNi alloy. A hypothetical BCC Al$_{0.3}$CoCrFeNi phase, with slightly increased volume (cf.~Table~\ref{tab:eos}), would exhibit even lower solution energies than FCC Al$_{0.3}$CoCrFeNi, and increasing its volume comparable to the Al-rich alloys of $12.6$~\AA$^3$/atom would reduce them further. This shows that, if stabilized, a BCC variant of the Al$_{0.3}$CoCrFeNi composition---or an FCC phase with an expanded lattice---would accommodate hydrogen more readily.

Overall, Fig.~\ref{fig:master_solution} demonstrates that hydrogen solution energies are governed primarily by Al content and ordering, secondarily by volume. At the same time, although directly correlated with changes in composition and material properties, the structural changes have only a modest effect once composition and volume are fixed in the simulations.

Finally, we examine whether magnetic effects could influence the energetic trend with hydrogen. Since magnetism is only implicitly considered in the GRACE potential, we conducted this analysis using DFT. To obtain an upper bound on the possible impact of magnetism, this analysis focused on the hydride formation energies of the two stable parent alloy phases, B2~Al$_3$CoCrFeNi and FCC~Al$_{0.3}$CoCrFeNi, as discussed in Sec.~\ref{sec:validation}. We evaluated these hydride formation energies using both spin-polarized and non-spin-polarized DFT calculations, with the latter representing the limiting case in which magnetic energies are suppressed. The non-spin-polarized results, shown in parentheses in Table~\ref{tab:eos_hydrides}, differ only slightly from the spin-polarized values. Importantly, both the absolute hydride formation energies and, crucially, the energy differences between competing hydrides remain essentially unchanged. This indicates that magnetic contributions play a negligible role in our case when comparing the relative stabilities of hydrides and therefore do not affect the trends in hydrogen affinity identified earlier. A detailed analysis of element-resolved local magnetic moments, including their pressure dependence in the parent alloys, is provided in the Supplemental Material.

\section{Conclusions}

We have investigated hydrogen uptake and hydride formation in Al-containing high-entropy alloys by combining high-pressure experiments, DFT, and a GRACE foundational potential. An important prerequisite for the hydrogen-uptake analysis was the validation of GRACE for hydrogen energetics in chemically complex transition-metal alloys. We have demonstrated that GRACE reliably reproduces DFT-level trends for hydrogen-free alloys, dilute-limit hydrogen solution energies, and hydride configurations. While small systematic offsets in absolute energies arise from the hydrogen reference chemical potential, relative energetics, volume changes, and stability are consistently captured.

Complementary high-pressure experiments conducted with DAC and LVP provide a direct benchmark for our predictions. At ambient temperature, FCC Al$_{0.3}$CoCrFeNi readily absorbs hydrogen and forms hydrides under elevated pressure, whereas B2 Al$_3$CoCrFeNi remains inert even at hydrogen pressures up to 50~GPa. The experimentally observed volume expansion and the qualitative differences in hydride formation behavior are in good agreement with the pressure–volume relations and hydride energetics obtained from DFT and GRACE. Although the calculations differ slightly in absolute volumes, they describe well the magnitude of the volume change upon hydriding and the contrasting hydrogen affinities of the two alloys, with the remaining absolute-volume offsets consistent with known PBE volume trends for transition metals and Al.

By systematically varying composition, chemical order, volume, and crystal structure, we have identified the dominant factors governing hydrogen solution energies in Al$_x$CoCrFeNi alloys. The aluminum content primarily controls hydrogen affinity. Increasing the Al concentration strongly destabilizes interstitial hydrogen through chemical effects. Changes in volume provide a secondary contribution: the larger equilibrium volumes of Al-rich alloys partially compensate for the chemical penalty but do not reverse it. Chemical ordering further suppresses hydrogen uptake by reducing the statistical occurrence of locally Al-poor environments, although low-energy hydrogen sites are not strictly forbidden even in the B2-ordered alloy. In contrast, the choice between FCC and BCC lattices has only a minor effect once composition and volume are fixed. For the equilibrium FCC and B2 reference phases, the relative hydride formation energies—and hence the stability between competing hydrides—are robust with respect to magnetic contributions.

Overall, our results demonstrate that hydrogen uptake in Al-containing high-entropy alloys is primarily governed by chemical composition, with chemical ordering providing a secondary effect, while the underlying crystal structure and magnetism play a comparatively minor role once composition and volume are fixed. The ability of GRACE to reproduce DFT energetics while enabling exhaustive sampling of hydrogen environments highlights its value for studying hydrogen behavior in complex alloys. This approach provides a practical route for composition-driven screening and design of hydrogen-resistant or hydrogen-affine multicomponent materials.

\section*{CRediT authorship contributions}

\textbf{Fritz K\"ormann} conceived the research, developed the methodology, performed the calculations, acquired funding, and wrote the original manuscript draft.
\textbf{Yuji Ikeda} contributed to the methodology and writing of the original draft and participated in manuscript review and editing.
\textbf{Konstantin Glazyrin}, \textbf{Kristina Spektor}, and \textbf{Kirill V. Yusenko} conceived the research, developed the experimental methodology, and acquired funding. Together with \textbf{Maxim Bykov}, \textbf{Shrikant Bhat}, and \textbf{Nikita Y. Gugin}, they performed the high-pressure experiments, and experimental analysis and participated in manuscript review and editing.
\textbf{Blazej Grabowski} contributed to the original draft and participated in manuscript review and editing.
\textbf{Anton Bochkarev}, \textbf{Yury Lysogorskiy}, and \textbf{Ralf Drautz} developed GRACE and the foundational potential, contributed to methodology, and participated in writing and editing the manuscript.
All authors discussed the results and approved the final manuscript.

\section*{Declaration of Generative AI and AI-assisted technologies in the writing process}

Grammarly was used to assist with sentence refinement for certain sections of this manuscript. The authors subsequently reviewed and edited the manuscript in its full extent  and take full responsibility for the final version of the publication.

\section*{Declaration of competing interest}

The authors declare that they have no known competing financial interests or personal relationships that influenced the work reported in this paper.

\section*{Acknowledgments}

Fritz K\"ormann acknowledges support by the Heisenberg Programme of the Deutsche Forschungsgemeinschaft (DFG, German Research Foundation), Project No. 541649719 and No. 569255040.
Yuji Ikeda acknowledges support from the DFG through project No. 519607530 (IK 125/1-1).
Blazej Grabowski acknowledges funding from the European Research Council (ERC) under the European Union’s Horizon Europe Research and Innovation Programme (Grant Agreement No.\ 101200433, project META-LEARN). Anton Bochkarev and Ralf Drautz acknowledge funding from the Deutsche Forschungsgemeinschaft (DFG, German Research Foundation) through SFB 1394 (project number 409476157).
Anton Bochkarev and Yury Lysogorskiy acknowledge high-performance computing resources provided by the Paderborn Center for Parallel Computing (PC$^2$) and the Elysium HPC cluster at Ruhr-University Bochum.
We acknowledge DESY (Hamburg, Germany), a member of the Helmholtz Association HGF, for the provision of experimental facilities.
Parts of this research were carried out at beamlines P02.2 and P61B with the support from the Federal Ministry of Education and Research, Germany (BMBF, grants no.: 05K16WC2 $\&$ 05K13WC2) and DESY (under proposal numbers I-20230202 and I-20220244). We are grateful to Dr.~Robert Farla for his support during P61B experiments.
Kristina Spektor thanks Per Mistenius for skillfully manufacturing the miniature press dies used for sample preparation.
We also thank the ID15B beamline (proposal number MA5925) at the European Synchrotron Radiation Facility, Grenoble, France, for providing us with the measurement time and technical support.

Funded by the European Union. Views and opinions expressed are however those of the author(s) only and do not necessarily reflect those of the European Union or the European Research Council Executive Agency. Neither the European Union nor the granting authority can be held responsible for them.

\section*{Data availability}

The datasets generated in this study are available from the corresponding author upon reasonable request.


\small
\begin{thebibliography}{10}
\expandafter\ifx\csname url\endcsname\relax
  \def\url#1{\texttt{#1}}\fi
\expandafter\ifx\csname urlprefix\endcsname\relax\def\urlprefix{URL }\fi
\expandafter\ifx\csname href\endcsname\relax
  \def\href#1#2{#2} \def\path#1{#1}\fi

\bibitem{marques2021review}
F.~Marques, M.~Balcerzak, F.~Winkelmann, G.~Zepon, M.~Felderhoff, Review and
  outlook on high-entropy alloys for hydrogen storage, Energy Environ. Sci.
  14~(10) (2021) 5191--5227.
\newblock \href {https://doi.org/https://doi.org/10.1039/D1EE01543E}
  {\path{doi:https://doi.org/10.1039/D1EE01543E}}.

\bibitem{LUO2024406}
L.~Luo, L.~Chen, L.~Li, S.~Liu, Y.~Li, C.~Li, L.~Li, J.~Cui, Y.~Li,
  High-entropy alloys for solid hydrogen storage: a review, Int. J. Hydrogen
  Energy 50 (2024) 406--430.
\newblock \href
  {https://doi.org/https://doi.org/10.1016/j.ijhydene.2023.07.146}
  {\path{doi:https://doi.org/10.1016/j.ijhydene.2023.07.146}}.

\bibitem{li2022hydrogen}
X.~Li, J.~Yin, J.~Zhang, Y.~Wang, X.~Song, Y.~Zhang, X.~Ren, Hydrogen
  embrittlement and failure mechanisms of multi-principal element alloys: A
  review, J. Mater. Sci. Technol. 122 (2022) 20--32.
\newblock \href {https://doi.org/https://doi.org/10.1016/j.jmst.2022.01.008}
  {\path{doi:https://doi.org/10.1016/j.jmst.2022.01.008}}.

\bibitem{YIN2023105306}
X.~Yin, X.~Liu, H.~Chen, S.~Chen, Hydrogen segregation by local chemical
  ordering structure in {CrCoN} medium-entropy alloys: A first principle study,
  Mater. Today Commun. 34 (2023) 105306.
\newblock \href {https://doi.org/https://doi.org/10.1016/j.mtcomm.2022.105306}
  {\path{doi:https://doi.org/10.1016/j.mtcomm.2022.105306}}.

\bibitem{li2010effect}
C.~Li, J.~Li, M.~Zhao, Q.~Jiang, Effect of aluminum contents on microstructure
  and properties of {Al$_x$CoCrFeNi} alloys, J. Alloys Compd. 504 (2010)
  S515--S518.
\newblock \href {https://doi.org/https://doi.org/10.1016/j.jallcom.2010.03.111}
  {\path{doi:https://doi.org/10.1016/j.jallcom.2010.03.111}}.

\bibitem{wang2014phases}
W.-R. Wang, W.-L. Wang, J.-W. Yeh, Phases, microstructure and mechanical
  properties of {Al$_x$CoCrFeNi} high-entropy alloys at elevated temperatures,
  J. Alloys Compd. 589 (2014) 143--152.
\newblock \href {https://doi.org/https://doi.org/10.1016/j.jallcom.2013.11.084}
  {\path{doi:https://doi.org/10.1016/j.jallcom.2013.11.084}}.

\bibitem{Wolverton2004}
C.~Wolverton, V.~Ozoli\ifmmode \mbox{\c{n}}\else
  \c{n}\fi{}\ifmmode~\check{s}\else \v{s}\fi{}, M.~Asta, Hydrogen in aluminum:
  First-principles calculations of structure and thermodynamics, Phys. Rev. B
  69 (2004) 144109.
\newblock \href {https://doi.org/10.1103/PhysRevB.69.144109}
  {\path{doi:10.1103/PhysRevB.69.144109}}.

\bibitem{glazyrin2024}
K.~Glazyrin, K.~Spektor, M.~Bykov, W.~Dong, J.-H.~Y. Yu, S.~Yang, J.-S.~L. Lee,
  S.~V. Divinski, M.~Hanfland, K.~V. Yusenko, High-entropy alloys and their
  affinity with hydrogen: From cantor to platinum group elements alloys, Adv.
  Sci. 11~(31) (2024) 2401741.
\newblock \href {https://doi.org/https://doi.org/10.1002/advs.202401741}
  {\path{doi:https://doi.org/10.1002/advs.202401741}}.

\bibitem{glazyrin_natcomm_accepted}
K.~Glazyrin, K.~Spektor, M.~Bykov, P.~H.~B. Brant~Carvalho, W.~Dong,
  F.~K{\"o}rmann, A.~Sano-Furukawa, T.~Hattori, D.~Beyer, M.~Sahlberg,
  Y.~Ikeda, J.-H. Yu, S.~Yang, J.-S. Lee, M.~Hanfland, B.~Grabowski,
  S.~Divinski, K.~V. Yusenko, Synthesis of high-entropy hydride from {Cantor}
  alloy (fcc–{CoCrFeNiMn}) at extreme conditions, Nat. Comm.{ in press}
  (2026).
\newblock \href {https://doi.org/10.1038/s41467-026-70483-3}
  {\path{doi:10.1038/s41467-026-70483-3}}.

\bibitem{qian2025hydrogen}
Y.~Qian, A.~Tamm, D.~Cereceda, S.~Kang, Hydrogen and water interactions with
  {CrMnFeCoNi} alloy from density functional theory calculations, Comput.
  Mater. Sci. 252 (2025) 113789.
\newblock \href
  {https://doi.org/https://doi.org/10.1016/j.commatsci.2025.113789}
  {\path{doi:https://doi.org/10.1016/j.commatsci.2025.113789}}.

\bibitem{wu2025hydrogen}
C.~Wu, Y.~Gong, C.~Liu, X.~Li, G.~Gizer, C.~Pistidda, F.~K{\"o}rmann, Y.~Ma,
  J.~Neugebauer, D.~Raabe, Hydrogen accommodation and its role in lattice
  symmetry in a {TiNbZr} medium-entropy alloy, Acta Mater. 288 (2025) 120852.
\newblock \href {https://doi.org/https://doi.org/10.1016/j.actamat.2025.120852}
  {\path{doi:https://doi.org/10.1016/j.actamat.2025.120852}}.

\bibitem{moore2022hydrogen}
C.~Moore, J.~Wilson, M.~Rushton, W.~Lee, J.~Astbury, S.~Middleburgh, Hydrogen
  accommodation in the {TiZrNbHfTa} high entropy alloy, Acta Mater. 229 (2022)
  117832.
\newblock \href {https://doi.org/https://doi.org/10.1016/j.actamat.2022.117832}
  {\path{doi:https://doi.org/10.1016/j.actamat.2022.117832}}.

\bibitem{ren2021hydrogen}
X.~Ren, P.~Shi, B.~Yao, L.~Wu, X.~Wu, Y.~Wang, Hydrogen solution in
  high-entropy alloys, Phys. Chem. Chem. Phys. 23~(48) (2021) 27185--27194.
\newblock \href {https://doi.org/10.1039/d1cp04151g}
  {\path{doi:10.1039/d1cp04151g}}.

\bibitem{jacobs2025practical}
R.~Jacobs, D.~Morgan, S.~Attarian, J.~Meng, C.~Shen, Z.~Wu, C.~Y. Xie, J.~H.
  Yang, N.~Artrith, B.~Blaiszik, et~al., A practical guide to machine learning
  interatomic potentials--status and future, Curr. Opin. Solid State Mater.
  Sci. 35 (2025) 101214.
\newblock \href {https://doi.org/10.1016/j.cossms.2025.101214}
  {\path{doi:10.1016/j.cossms.2025.101214}}.

\bibitem{PhysRevX.14.021036}
A.~Bochkarev, Y.~Lysogorskiy, R.~Drautz, Graph atomic cluster expansion for
  semilocal interactions beyond equivariant message passing, Phys. Rev. X 14
  (2024) 021036.
\newblock \href {https://doi.org/10.1103/PhysRevX.14.021036}
  {\path{doi:10.1103/PhysRevX.14.021036}}.

\bibitem{lysogorskiy2025graphatomicclusterexpansion}
Y.~Lysogorskiy, A.~Bochkarev, R.~Drautz, Graph atomic cluster expansion for
  foundational machine learning interatomic potentials, npj Comput. Mater.
  (2026).
\newblock \href {https://doi.org/10.1038/s41524-026-01979-1}
  {\path{doi:10.1038/s41524-026-01979-1}}.

\bibitem{zhang2025lattice}
J.~Zhang, X.~Xu, F.~K{\"o}rmann, W.~Yin, X.~Zhang, C.~Gadelmeier, U.~Glatzel,
  B.~Grabowski, R.~Li, G.~Liu, et~al., Lattice distortions and non-sluggish
  diffusion in bcc refractory high entropy alloys, Acta Mater. (2025)
  121283\href {https://doi.org/10.1016/j.actamat.2025.121283}
  {\path{doi:10.1016/j.actamat.2025.121283}}.

\bibitem{kresse1993}
G.~Kresse, J.~Hafner, Ab initio molecular dynamics for liquid metals, Phys.
  Rev. B 47~(1) (1993) 558.
\newblock \href {https://doi.org/https://doi.org/10.1103/PhysRevB.47.558}
  {\path{doi:https://doi.org/10.1103/PhysRevB.47.558}}.

\bibitem{kresse1994}
G.~Kresse, J.~Hafner, Ab initio molecular-dynamics simulation of the
  liquid-metal--amorphous-semiconductor transition in germanium, Phys. Rev. B
  49~(20) (1994) 14251.
\newblock \href {https://doi.org/https://doi.org/10.1103/PhysRevB.49.14251}
  {\path{doi:https://doi.org/10.1103/PhysRevB.49.14251}}.

\bibitem{kresse1996}
G.~Kresse, J.~Furthm{\"u}ller, Efficiency of ab-initio total energy
  calculations for metals and semiconductors using a plane-wave basis set,
  Comput. Mater. Sci. 6~(1) (1996) 15--50.
\newblock \href {https://doi.org/https://doi.org/10.1016/0927-0256(96)00008-0}
  {\path{doi:https://doi.org/10.1016/0927-0256(96)00008-0}}.

\bibitem{blochl1994}
P.~E. Bl{\"o}chl, Projector augmented-wave method, Phys. Rev. B 50~(24) (1994)
  17953.
\newblock \href {https://doi.org/https://doi.org/10.1103/PhysRevB.50.17953}
  {\path{doi:https://doi.org/10.1103/PhysRevB.50.17953}}.

\bibitem{perdew1996}
J.~P. Perdew, K.~Burke, M.~Ernzerhof, Generalized gradient approximation made
  simple, Phys. Rev. Lett. 77~(18) (1996) 3865.
\newblock \href {https://doi.org/https://doi.org/10.1103/PhysRevLett.77.3865}
  {\path{doi:https://doi.org/10.1103/PhysRevLett.77.3865}}.

\bibitem{zunger1990special}
A.~Zunger, S.-H. Wei, L.~G. Ferreira, J.~E. Bernard, Special quasirandom
  structures, Phys. Rev. Lett. 65~(3) (1990) 353.
\newblock \href {https://doi.org/10.1103/PhysRevLett.65.353}
  {\path{doi:10.1103/PhysRevLett.65.353}}.

\bibitem{aangqvist2019icet}
M.~{\AA}ngqvist, W.~A. Mu{\~n}oz, J.~M. Rahm, E.~Fransson, C.~Durniak,
  P.~Rozyczko, T.~H. Rod, P.~Erhart, {ICET}--a {Python} library for
  constructing and sampling alloy cluster expansions, Adv. Theory Simul. 2~(7)
  (2019) 1900015.
\newblock \href {https://doi.org/https://doi.org/10.1002/adts.201900015}
  {\path{doi:https://doi.org/10.1002/adts.201900015}}.

\bibitem{barroso2024open}
L.~Barroso-Luque, M.~Shuaibi, X.~Fu, B.~M. Wood, M.~Dzamba, M.~Gao, A.~Rizvi,
  C.~L. Zitnick, Z.~W. Ulissi, Open materials 2024 ({OMAT24}) inorganic
  materials dataset and models (2024).
\newblock \href {http://arxiv.org/abs/2410.12771} {\path{arXiv:2410.12771}}.

\bibitem{pa2025information}
A.~PA~Subramanyam, D.~Perez, Information-entropy-driven generation of
  material-agnostic datasets for machine-learning interatomic potentials, npj
  Comput. Mater. 11~(1) (2025) 218.
\newblock \href {https://doi.org/10.1038/s41524-025-01602-9}
  {\path{doi:10.1038/s41524-025-01602-9}}.

\bibitem{maxEntropy_foundation_submitted}
A.~Bochkarev, Y.~Lysogorskiy, A.~Subramanyam, R.~Drautz, D.~Perez, Exploring
  the extremes: atomic basis for multi-elemental materials science under
  complex thermodynamic conditions, arXiv preprint (2026).
\newblock \href {https://doi.org/10.48550/arXiv.2602.23489}
  {\path{doi:10.48550/arXiv.2602.23489}}.

\bibitem{larsen2017atomic}
A.~H. Larsen, J.~J. Mortensen, J.~Blomqvist, I.~E. Castelli, R.~Christensen,
  M.~Du{\l}ak, J.~Friis, M.~N. Groves, B.~Hammer, C.~Hargus, et~al., The atomic
  simulation environment—a {Python} library for working with atoms, J. Phys.
  Condens. Matter 29~(27) (2017) 273002.
\newblock \href {https://doi.org/10.1088/1361-648X/aa680e}
  {\path{doi:10.1088/1361-648X/aa680e}}.

\bibitem{Yusenko2018}
K.~V. Yusenko, S.~Riva, W.~A. Crichton, K.~Spektor, E.~Bykova, A.~Pakhomova,
  A.~Tudball, I.~Kupenko, A.~Rohrbach, S.~Klemme, et~al., High-pressure
  high-temperature tailoring of high entropy alloys for extreme environments,
  J. Alloys Compd. 738 (2018) 491--500.
\newblock \href {https://doi.org/https://doi.org/10.1016/j.jallcom.2017.12.216}
  {\path{doi:https://doi.org/10.1016/j.jallcom.2017.12.216}}.

\bibitem{Liermann2015}
H.-P. Liermann, Z.~Kon{\^o}pkov{\'a}, W.~Morgenroth, K.~Glazyrin,
  J.~Bednar{\v{c}}ik, E.~McBride, S.~Petitgirard, J.~Delitz, M.~Wendt,
  Y.~Bican, et~al., The extreme conditions beamline {P02.2} and the extreme
  conditions science infrastructure at {PETRA III}, J. Synchrotron Radiat.
  22~(4) (2015) 908--924.
\newblock \href {https://doi.org/10.1107/S1600577515005937}
  {\path{doi:10.1107/S1600577515005937}}.

\bibitem{Fei2007}
Y.~Fei, A.~Ricolleau, M.~Frank, K.~Mibe, G.~Shen, V.~Prakapenka, Toward an
  internally consistent pressure scale, Proc. Natl. Acad. Sci. U.S.A. 104~(22)
  (2007) 9182--9186.
\newblock \href {https://doi.org/10.1073/pnas.0609013104}
  {\path{doi:10.1073/pnas.0609013104}}.

\bibitem{Prescher2015}
C.~Prescher, V.~B. Prakapenka, Dioptas: a program for reduction of
  two-dimensional {X-ray} diffraction data and data exploration, High Press.
  Res. 35~(3) (2015) 223--230.
\newblock \href {https://doi.org/10.1080/08957959.2015.1059835}
  {\path{doi:10.1080/08957959.2015.1059835}}.

\bibitem{Coelho2018}
A.~A. Coelho, Topas and topas-academic: an optimization program integrating
  computer algebra and crystallographic objects written in c++, J. Appl.
  Crystallogr. 51~(1) (2018) 210--218.
\newblock \href {https://doi.org/10.1107/S1600576718000183}
  {\path{doi:10.1107/S1600576718000183}}.

\bibitem{GonzalezPlatas2016}
J.~Gonzalez-Platas, M.~Alvaro, F.~Nestola, R.~Angel, {EosFit7-GUI}: a new
  graphical user interface for equation of state calculations, analyses and
  teaching, J. Appl. Crystallogr. 49~(4) (2016) 1377--1382.
\newblock \href {https://doi.org/10.1107/S1600576716008050}
  {\path{doi:10.1107/S1600576716008050}}.

\bibitem{Farla2022}
R.~Farla, S.~Bhat, S.~Sonntag, A.~Chanyshev, S.~Ma, T.~Ishii, Z.~Liu,
  A.~N{\'e}ri, N.~Nishiyama, G.~A. Faria, et~al., Extreme conditions research
  using the large-volume press at the {P61B} endstation, {PETRA III}, J.
  Synchrotron Radiat. 29~(2) (2022) 409--423.
\newblock \href {https://doi.org/10.1107/S1600577522001047}
  {\path{doi:10.1107/S1600577522001047}}.

\bibitem{Nylen2009}
J.~Nyl{\'e}n, T.~Sato, E.~Soignard, J.~L. Yarger, E.~Stoyanov,
  U.~H{\"a}ussermann, Thermal decomposition of ammonia borane at high
  pressures, J. Chem. Phys. 131~(10) (2009) 104506.
\newblock \href {https://doi.org/10.1063/1.3230973}
  {\path{doi:10.1063/1.3230973}}.

\bibitem{Vekilova2023}
O.~Y. Vekilova, D.~C. Beyer, S.~Bhat, R.~Farla, V.~Baran, S.~I. Simak,
  H.~Kohlmann, U.~Haussermann, K.~Spektor, Formation and polymorphism of
  semiconducting {K\textsubscript{2}SiH\textsubscript{6}} and strategy for
  metallization, Inorganic Chemistry 62~(21) (2023) 8093--8100.
\newblock \href {https://doi.org/10.1021/acs.inorgchem.2c04370}
  {\path{doi:10.1021/acs.inorgchem.2c04370}}.

\bibitem{matsui2012simultaneous}
M.~Matsui, Y.~Higo, Y.~Okamoto, T.~Irifune, K.-I. Funakoshi, Simultaneous sound
  velocity and density measurements of nacl at high temperatures and pressures:
  Application as a primary pressure standard, Am. Mineral. 97~(10) (2012)
  1670--1675.
\newblock \href {https://doi.org/10.2138/am.2012.4136}
  {\path{doi:10.2138/am.2012.4136}}.

\bibitem{Seto2010}
Y.~Seto, Development of a software suite on {X-ray} diffraction experiments,
  Rev. High Press. Sci. Technol. 20 (2010) 269--276.
\newblock \href {https://doi.org/10.4131/jshpreview.20.269}
  {\path{doi:10.4131/jshpreview.20.269}}.

\bibitem{smekhova2022driven}
A.~Smekhova, A.~Kuzmin, K.~Siemensmeyer, C.~Luo, K.~Chen, F.~Radu, E.~Weschke,
  U.~Reinholz, A.~G. Buzanich, K.~V. Yusenko, Al-driven peculiarities of local
  coordination and magnetic properties in single-phase {Al$_x$CrFeCoNi}
  high-entropy alloys, Nano Research 15~(6) (2022) 4845--4858.
\newblock \href {https://doi.org/10.1007/s12274-021-3704-5}
  {\path{doi:10.1007/s12274-021-3704-5}}.

\bibitem{PhysRevB.35.1945}
P.~Vinet, J.~R. Smith, J.~Ferrante, J.~H. Rose, Temperature effects on the
  universal equation of state of solids, Phys. Rev. B 35 (1987) 1945--1953.
\newblock \href {https://doi.org/10.1103/PhysRevB.35.1945}
  {\path{doi:10.1103/PhysRevB.35.1945}}.

\bibitem{zhao2023investigation}
Y.~Zhao, Z.~Chen, K.~Yan, W.~Le, S.~Naseem, H.~Zhang, L.~Yang, Investigation on
  microstructure, superior tensile property and its mechanism in
  {Al$_{0.3}$CoCrFeNi} high-entropy alloy via thermo-mechanical processing,
  Mater. Sci. Eng. A 866 (2023) 144690.
\newblock \href {https://doi.org/https://doi.org/10.1016/j.msea.2023.144690}
  {\path{doi:https://doi.org/10.1016/j.msea.2023.144690}}.

\bibitem{somenkov1987crystal}
V.~Somenkov, V.~Glazkov, A.~Irodova, S.~S. Shilstein, Crystal structure and
  volume effects in the hydrides of d metals, J. Less-Common Met. 129 (1987)
  171--180.
\newblock \href {https://doi.org/10.1016/0022-5088(87)90045-2}
  {\path{doi:10.1016/0022-5088(87)90045-2}}.

\bibitem{fukai2005metal}
Y.~Fukai, The metal-hydrogen system: basic bulk properties, Springer, 2005.

\bibitem{WANG2019468}
Z.~Wang, Q.~Wu, W.~Zhou, F.~He, C.~Yu, D.~Lin, J.~Wang, C.~Liu, Quantitative
  determination of the lattice constant in high entropy alloys, Scripta
  Materialia 162 (2019) 468--471.
\newblock \href
  {https://doi.org/https://doi.org/10.1016/j.scriptamat.2018.12.022}
  {\path{doi:https://doi.org/10.1016/j.scriptamat.2018.12.022}}.

\bibitem{PhysRevB.87.214102}
L.~Schimka, R.~Gaudoin, J.~c.~v. Klime\ifmmode~\check{s}\else \v{s}\fi{},
  M.~Marsman, G.~Kresse, Lattice constants and cohesive energies of alkali,
  alkaline-earth, and transition metals: Random phase approximation and density
  functional theory results, Phys. Rev. B 87 (2013) 214102.
\newblock \href {https://doi.org/10.1103/PhysRevB.87.214102}
  {\path{doi:10.1103/PhysRevB.87.214102}}.

\bibitem{Lejaeghere_2006}
K.~Lejaeghere, G.~Bihlmayer, T.~Björkman, P.~Blaha, S.~Blügel, V.~Blum,
  D.~Caliste, I.~E. Castelli, S.~J. Clark, A.~D. Corso, S.~de~Gironcoli,
  T.~Deutsch, J.~K. Dewhurst, I.~D. Marco, C.~Draxl, M.~Dułak, O.~Eriksson,
  J.~A. Flores-Livas, K.~F. Garrity, L.~Genovese, P.~Giannozzi, M.~Giantomassi,
  S.~Goedecker, X.~Gonze, O.~Grånäs, E.~K.~U. Gross, A.~Gulans, F.~Gygi,
  D.~R. Hamann, P.~J. Hasnip, N.~A.~W. Holzwarth, D.~Iuşan, D.~B. Jochym,
  F.~Jollet, D.~Jones, G.~Kresse, K.~Koepernik, E.~Küçükbenli, Y.~O.
  Kvashnin, I.~L.~M. Locht, S.~Lubeck, M.~Marsman, N.~Marzari, U.~Nitzsche,
  L.~Nordström, T.~Ozaki, L.~Paulatto, C.~J. Pickard, W.~Poelmans, M.~I.~J.
  Probert, K.~Refson, M.~Richter, G.-M. Rignanese, S.~Saha, M.~Scheffler,
  M.~Schlipf, K.~Schwarz, S.~Sharma, F.~Tavazza, P.~Thunström, A.~Tkatchenko,
  M.~Torrent, D.~Vanderbilt, M.~J. van Setten, V.~V. Speybroeck, J.~M. Wills,
  J.~R. Yates, G.-X. Zhang, S.~Cottenier, Reproducibility in density functional
  theory calculations of solids, Science 351~(6280) (2016) aad3000.
\newblock \href {https://doi.org/10.1126/science.aad3000}
  {\path{doi:10.1126/science.aad3000}}.

\bibitem{PhysRevB.79.085104}
P.~Haas, F.~Tran, P.~Blaha, Calculation of the lattice constant of solids with
  semilocal functionals, Phys. Rev. B 79 (2009) 085104.
\newblock \href {https://doi.org/10.1103/PhysRevB.79.085104}
  {\path{doi:10.1103/PhysRevB.79.085104}}.

\end{thebibliography}
\end{document}